\documentclass[conference]{IEEEtran}

\usepackage[utf8]{inputenc} 
\usepackage[T1]{fontenc}
\usepackage{url}
\usepackage{ifthen}
\usepackage{cite}
\usepackage{amsfonts, amsthm, amssymb, dsfont}
\usepackage[cmex10]{amsmath} 
\usepackage[a4paper, top=0.84in, bottom=1.8in, left=1.4cm, right=1.4cm]{geometry}

\makeatletter
\newcommand*\myeqnum{\refstepcounter{equation}\tag{\theequation}}
\makeatother

\IEEEoverridecommandlockouts

\newtheorem{theorem}{Theorem}

\begin{document}
\title{Dual-Domain Exponent of Maximum\\ Mutual Information Decoding} 

\author{%
  \IEEEauthorblockN{Seyed AmirPouya Moeini}
  \IEEEauthorblockA{University of Cambridge\\
    \texttt{sam297@cam.ac.uk}}
  \and
  \IEEEauthorblockN{Albert Guill\'en i F\`abregas}
  \IEEEauthorblockA{University of Cambridge\\
  Universitat Polit\`ecnica de Catalunya\\
    \texttt{guillen@ieee.org}}
  \thanks{This work was supported in part by the European Research Council under Grants 725411 and 101142747, and in part by the Spanish Ministry of Economy and Competitiveness under Grant PID2020-116683GB-C22.}
}

\newcommand{\bs}[1]{\boldsymbol{#1}}
\newcommand{\mcl}[1]{\mathcal{#1}}

\maketitle

\begin{abstract}
This paper provides a dual domain derivation of the error exponent of maximum mutual information (MMI) decoding with constant composition codes, showing it coincides with that of maximum likelihood decoding for discrete memoryless channels. The analysis is further extended to joint source-channel coding, demonstrating that the generalized MMI decoder achieves the same random coding error exponent as the maximum a posteriori decoder.
\end{abstract}

\section{Introduction}

The random coding error exponent is defined as the asymptotic limit of the negative normalized logarithm of the expected error probability, given by
\begin{align}
    E_{\mathrm{r}}(R) \triangleq \lim _{n \rightarrow \infty} -\frac{1}{n} \log \bar{p}_e,
\end{align}
where $R$ is the coding rate, and $\bar{p}_e$ denotes the average error probability over all codebooks in the ensemble.

When messages are equiprobable, Csiszár and Körner \cite[Theorem 10.2]{csiszar2011information} (see also Gallager \cite[Section~6]{gallager_fixed}) showed that, for discrete memoryless channels (DMCs) and under constant-composition coding, the maximum mutual information (MMI) decoder achieves the same random coding error exponent as the maximum likelihood (ML) decoder.
The MMI decoder is a suboptimal mismatched decoder that does not know the channel transition probability and yet attains the same error exponent as ML decoding. 
This notion of optimality with respect to the random coding error exponent is commonly termed as \emph{universality}.

Csiszár and Körner's expression results in the minimization over joint distributions subject to specific constraints. 
These expressions are typically referred to as primal domain, and are typically obtained using the method of types \cite[Chapter~2]{csiszar2011information}. 
Instead, dual expressions are commonly formulated as maximizations over auxiliary parameters, with key approaches for deriving random coding bounds in the dual domain being those of Gallager \cite[Section~5.6]{gallager} and Poltyrev \cite{poltyrev}. 
A dual-domain derivation offers notable advantages, such as enabling achievable exponents to be obtained for any parameter choice and providing a natural extension to arbitrary alphabets.

Existing results of the MMI decoder have been limited to the primal domain, as its multi-letter nature makes a dual-domain analysis difficult.
In this work, we revisit the universality of the MMI decoder in channel coding by deriving its random coding error exponent  in the dual domain. 
We then extend our analysis to joint source-channel coding (JSCC), where messages are not equiprobable, by addressing the generalized MMI decoder introduced by Csiszár \cite{csiszar1980}.
To the best of our knowledge, such dual-domain derivations for the MMI decoder have not been previously appeared in the literature. 

\section{Notations}

In this paper, scalar random variables are denoted by uppercase letters, their sample values by lowercase letters, and their alphabets by calligraphic letters. Random vectors are represented in boldface.
For two positive sequences $f_n$ and $g_n$, we write $f_n \doteq g_n$ if $\lim _{n \rightarrow \infty} \frac{1}{n} \log \frac{f_n}{g_n}=0$, and we write $f_n \, \dot{\leq}\,\, g_n$ if $\lim \sup _{n \rightarrow \infty} \frac{1}{n} \log \frac{f_n}{g_n} \leq 0$.

%
%
The type of a sequence $\bs{x}=\left(x_1, \ldots, x_n\right) \in \mcl{X}^n$ is its empirical distribution, defined by
\begin{equation}
    \hat{P}_{\bs{x}}(x) \triangleq \frac{1}{n} \sum_{i=1}^n \mathds{1}\left\{{x}_i=x\right\} .
\end{equation}
Similarly, the joint type and conditional types of a pair of sequences are denoted by $\hat{P}_{\bs{xy}}$, $\hat{P}_{\bs{y}|\bs{x}}$, and $\hat{P}_{\bs{x}|\bs{y}}$, and are defined in an analogous manner.
Throughout this paper, we use $\hat{P}_{\bs{x}}(\bs{x})$ for convenience to represent
$\hat{P}_{\bs{x}}(\bs{x}) = \prod_{i=1}^n \hat{P}_{\bs{x}}(x_i)$.
Observe that $\hat{P}_{\bs{x}}(\bs{x}) = \exp(-nH(\hat{P}_{\bs{x}}))$,
where $H(P)$ represents the entropy of the distribution $P$.
The set of all probability distributions on an alphabet $\mcl{X}$ is denoted by $\mcl{P}(\mcl{X})$, while $\mcl{P}_n(\mcl{X})$ represents the set of empirical distributions (types) for vectors in $\mcl{X}^n$.
It is shown in \cite[Lemma~2.2]{csiszar2011information} that the total number of types grows polynomially with $n$, which implies $|\mcl{P}_n(\mcl{X})| \doteq 1$.
For $P_X \in \mcl{P}_n(\mcl{X})$, the type class $\mcl{T}^n(P_X)$ consists of all sequences in $\mcl{X}^n$ with type $P_X$.
It is shown in \cite[Lemma~2.3]{csiszar2011information} that $|\mcl{T}^n(P_X)| \doteq \exp(nH(P_X))$.
Under the constant-composition random coding ensemble with input distribution $Q_X$, the codewords are independently drawn from the codeword distribution
\begin{align}
    P_{\bs{X}}(\bs{x})=\frac{1}{\left|\mathcal{T}^n(Q_X)\right|} \, \mathds{1}\left\{\bs{x} \in \mathcal{T}^n(Q_X)\right\}. \label{eq-cc-defn}
\end{align}

\section{Channel Coding}

We study communication over a DMC $W$. A codebook $\mcl{C}_n$ consists of $M$ codewords $\mcl{C}_n = \{\bs{x}(1), \ldots, \bs{x}(M )\}$, where each codeword $\bs{x}(m)\in\mcl{X}^n$ for $m \in \{1, \ldots, M \}$. 
The code rate $R$ is defined as $R = \frac{1}{n}\log M$, measured in nats per channel use.
A message $m$ is selected equiprobably from $\{1, \ldots, M \}$, and the corresponding codeword $\bs{x}(m)$ is sent through the channel. 
The channel produces an output sequence $\bs{y} = (y_1, \ldots, y_n) \in \mcl{Y}^n$ based on the conditional probability $W^n(\bs{y} | \bs{x}) = \prod_{i=1}^n W(y_i | x_i)$.
In \cite[Chapter 10]{csiszar2011information}, Csiszár and Körner used the method of types \cite[Chapter~2]{csiszar2011information} 
to show that the MMI decoder achieves the same random coding error exponent as the ML decoder.
In this section, we redrive this result in the dual domain.

Given the channel output $\bs{y}$, the MMI decoder selects the message $\hat{m}$ that maximizes the mutual information induced by the joint type between $\bs{y}$ and each of the codewords,
\begin{align}
     \hat{m} = \underset{m \in \{1,\ldots, M \}}{\operatorname{argmax}} I(\hat{P}_{\bs{x}(m) \bs{y}}),
\end{align}
where $ \hat{P}_{\bs{x}(m) \bs{y}}$ denotes the joint type of $\bs{x}(m)$ and $\bs{y}$.
This decoder is equivalent to selecting the message that maximizes the conditional type.
To see this, observe that
\begin{align}
    \begin{split}
        &\underset{m \in \{1,\ldots, M \}}{\operatorname{argmax}} I(\hat{P}_{\bs{x}(m) \bs{y}}) \label{eq-equality-py-0}\\
        &\hspace{1cm}= \underset{m \in \{1,\ldots, M \}}{\operatorname{argmax}}  \exp\left(nI(\hat{P}_{\bs{x}(m) \bs{y}})\right)\\
    \end{split}\\
    &\hspace{1.cm}= \underset{m \in \{1,\ldots, M \}}{\operatorname{argmax}}  \exp\left(n\left[H(\hat{P}_{\bs{y}})-H(\hat{P}_{\bs{y}|\bs{x}(m)})\right]\right)\\
    &\hspace{1.cm}= \underset{m \in \{1,\ldots, M \}}{\operatorname{argmax}}  \exp\left(-nH(\hat{P}_{\bs{y}|\bs{x}(m)})\right) \label{eq-equality-py}.
\end{align}
The equality in \eqref{eq-equality-py} follows from the fact that $H(\hat{P}_{\bs{y}})$ acts as a constant here, as it does not depend on $m$.
Furthermore, the conditional entropy $H(\hat{P}_{\bs{y}|\bs{x}(m)})$ can be expressed as
\begin{align}
    -nH(\hat{P}_{\bs{y}|\bs{x}(m)}) 
    &= \sum_{(a, b)} n\hat{P}_{\bs{x}(m) \bs{y}}(a, b) \log \hat{P}_{\bs{y}|\bs{x}(m)}(b|a)\\
    &= \log \prod_{(a, b)}\left[\hat{P}_{\bs{y}|\bs{x}(m)}(b|a) \right]^{n \hat{P}_{\bs{x}(m) \bs{y}}(a, b)}\\
    &= \log \prod_{i=1}^n  \hat{P}_{\bs{y}|\bs{x}(m)}(y_i|x_{m,i}).
\end{align}
Thus, we conclude that this decoder selects the message $\hat{m}$ as
\begin{align}
    \hat{m} = \underset{m \in \{1,\ldots, M \}}{\operatorname{argmax}} \prod_{i=1}^n  \hat{P}_{\bs{y}|\bs{x}(m)}(y_i|x_{m,i}). \label{eq-equality-py-last}
\end{align}
This form of the MMI decoder is more amenable for dual-domain derivations. 

\begin{theorem}
    For a DMC $W$ and rate $R$, under constant-composition coding with $Q_X$, we have
    \begin{align}
        \bar{p}_e^{\mathrm{\,\, mmi}}
        \,\, &\dot{\leq} \,\, \min_{0 \leq \rho \leq 1} \exp\left(-n \left[E_0^{\mathrm{ml}}(Q_X, \rho) - \rho R\right]\right),
    \end{align}
    where
    \begin{align}
        \begin{split}
            &E_0^{\mathrm{ml}}(Q_X, \rho)\\
            &= \sup_{r(.)} -\log \sum_y \left(\sum_x Q_X(x)W(y|x)^{\frac{1}{1+\rho}}e^{r(x)-\phi_r}\right)^{1+\rho}
        \end{split}
    \end{align}
    and $\phi_r = \sum_x Q_X(x)r(x)$. 
\end{theorem}

\begin{proof}
    We begin with the random coding union (RCU) bound, presented in \cite[Theorem 1]{6483461}, which states that for a generic decoding metric $q$, we have
    \begin{align}
        \begin{split}
	    &\bar{p}_e^{{\, q}} 
            \hspace{0cm}\doteq \mathbb{E}\!\left[\min\left\{1,  M \, \mathbb{P}\big[q(\bar{\bs{X}},\! \bs{Y})  \geq q(\bs{X},\! \bs{Y}) \big| \bs{X}, \bs{Y}\big]\right\}\right],
        \end{split}
    \end{align}
     where $\bar{p}_e^{{\,q}}$ denotes the average error probability under the decoding metric $q$.
    For the MMI decoder, we get
    \begin{align}
        \begin{split}
            \bar{p}_e^{\mathrm{\,\, mmi}} 
            &\doteq \mathbb{E}\left[\min\left\{1, M \, \mathbb{P}\left[\frac{I(\hat{P}_{\bs{\bar{X}Y}})}{I(\hat{P}_{\bs{XY}})} \geq 1 \, \big| \, \bs{X}, \bs{Y}\right]\right\}\right]
        \end{split} \\
        \begin{split}
            &= \, \mathbb{E}\Bigg[\min\biggl\{1, M \, \mathbb{P}\biggl[\frac{\hat{P}_{\bs{Y} | \bar{\bs{X}}}(\bs{Y} | \bs{\bar{X}})} { \hat{P}_{\bs{Y} | \bs{X}}(\bs{Y} | \bs{X})} \geq 1 \,\Big|\, \bs{X}, \bs{Y}\biggr]\biggr\}\Bigg].
        \end{split}
    \end{align}
    Given that $\min\{1, x\} \leq x^{\rho}$ for $x \geq 0$ and $\rho \in [0, 1]$, we can upper bound the expression for any $\rho \in [0, 1]$ as follows
    \begin{align}
        \bar{p}_e^{\mathrm{\,\, mmi}}
        &\!\leq \mathbb{E}\!\left[M ^{\rho} \mathbb{P}\!\left[\frac{\prod_{i=1}^n \hat{P}_{\bs{Y} | \bar{\bs{X}}}(Y_i | \bar{X}_i)}{\prod_{i=1}^n \hat{P}_{\bs{Y} | \bs{X}}(Y_i | X_i)} \! \geq 1 \,\bigg|\, \bs{X}, \bs{Y}\right]^{\rho}\right] \\
        &\!\leq \mathbb{E}\!\left[M ^{\rho} \left( \frac{\mathbb{E}\left[\prod_{i=1}^n \hat{P}_{\bs{Y} | \bar{\bs{X}}}(Y_i | \bar{X}_i) \,\big| \, \bs{Y}\right] }{\prod_{i=1}^n \hat{P}_{\bs{Y} | \bs{X}}(Y_i | X_i)}\right)^{\rho}\right], \label{eq-markov-up-1}
    \end{align}
    where in \eqref{eq-markov-up-1} we applied Markov's inequality. 
    Observe that, unlike standard dual-domain derivations, no tilting parameter $s$ is introduced.
	For a given $\bs{y}$,  \cite[Lemma~2.1]{poltyrev} allows us to rewrite the expectation in the numerator of \eqref{eq-markov-up-1} as if the distribution were i.i.d. with $Q_X$, as follows
    \begin{align}
	\mathbb{E}\left[\hat{P}_{\bs{y} | \bar{\bs{X}}}\left(\bs{y} | \bar{\bs{X}}\right)\right] 
        & \,\, \dot{\leq} \!\!\!\!\! \sum_{\bar{\bs{x}} \in \mcl{T}^n(Q_X)} \prod_{i=1}^n Q_X(\bar{x}_i) \hat{P}_{\bs{y} | \bar{\bs{x}}}\left(y_i | \bar{x}_i\right). \label{eq-pol0}
    \end{align}
    Recall that $Q_X = \hat{P}_{\bs{\bar{x}}}$, and note that for $a \in \mcl{X}$ and $b \in \mcl{Y}$, we have $\hat{P}_{\bs{\bar{x}}}(a)\hat{P}_{\bs{y}|\bs{\bar{x}}}(b|a) = \hat{P}_{\bs{y}}(b)\hat{P}_{\bs{\bar{x}}|\bs{y}}(a|b)$. Thus, the RHS of \eqref{eq-pol0} can be rewritten as
    \begin{align}
        \begin{split}
            &\mathbb{E}\left[\prod_{i=1}^n \hat{P}_{\bs{y} | \bar{\bs{X}}}\left({y}_i | \bar{{X}}_i\right)\right] \\
            &\hspace{0.65cm}\,\, \dot{\leq} \,\, \hat{P}_{\bs{y}}(\bs{y}) \!\!\! \sum_{\bar{\bs{x}} \in \mcl{T}^n(Q_X)} \prod_{i=1}^n \hat{P}_{\bar{\bs{x}}|\bs{y}}\left(\bar{x}_i|y_i\right) \label{eq-pol1}
        \end{split}\\
        &\hspace{0.65cm} = \hat{P}_{\bs{y}}(\bs{y})  \!\!\!\!\!\!\!\! \sum_{\substack{\bar{P}_{\bar{X}Y}:\\\bar{P}_{\bar{X}}=Q_X, \bar{P}_Y = \hat{P}_{\bs{y}}}} \!\!\!\sum_{\bar{\bs{x}}: (\bar{\bs{x}}, \bs{y})\in \mcl{T}^n(\bar{P})} \prod_{i=1}^n \bar{P}_{\bar{X}|Y}(\bar{x}_i|y_i) \label{eq-cond-entropy} \\
        &\hspace{0.65cm} \doteq \hat{P}_{\bs{y}}(\bs{y}) \!\!\!\!\!\!\!\! \sum_{\substack{\bar{P}_{\bar{X}Y}:\\\bar{P}_{\bar{X}}=Q_X, \bar{P}_Y = \hat{P}_{\bs{y}}}} 1 \label{eq-one-canc}\\
        &\hspace{0.65cm} \doteq \hat{P}_{\bs{y}}(\bs{y}),  \label{eq-pol2}
    \end{align}
    where $\bar{P}_{\bar{X}|Y}$ in \eqref{eq-cond-entropy} is the conditional distribution of $\bar{X}$ given $Y$ induced by the joint distribution $\bar{P}_{\bar{X}Y}$.
    The $1$ in \eqref{eq-one-canc} appears because the product $\prod_{i=1}^n \bar{P}_{\bar{X}|Y}(\bar{x}_i|y_i)$ simplifies to $\exp(-nH_{\bar{P}}(\bar{X}|Y))$, while the size of the set $\{\bar{\bs{x}} : (\bar{\bs{x}}, \bs{y}) \in \mcl{T}^n(\bar{P})\} \doteq \exp(nH_{\bar{P}}(\bar{X}|Y))$, and these terms cancel each other.
    Substituting this result into \eqref{eq-markov-up-1}, we obtain
    \begin{align}
	\bar{p}_e^{\mathrm{\,\, mmi}}\,
        &\, \dot{\leq} \,\,\mathbb{E}\left[M ^{\rho} \, \left( \frac{ \prod_{i=1}^n \hat{P}_{\bs{Y}}(Y_i)  }{\prod_{i=1}^n \hat{P}_{\bs{Y} | \bs{X}}(Y_i | X_i)}\right)^{\rho}\right]\\
        &= \mathbb{E}\left[M ^{\rho} \, \left( \frac{ \prod_{i=1}^n \hat{P}_{\bs{X}}(X_i)  }{\prod_{i=1}^n \hat{P}_{\bs{X}|\bs{Y}}(X_i|Y_i)}\right)^{\rho}\right].
        \label{eq:pxy_div}
    \end{align}
    For a given pair $(\bs{x}, \bs{y})$, the conditional type can be written as
    \begin{align}
        \prod_{i=1}^n \hat{P}_{\bs{x}|\bs{y}}(x_i|y_i) &= \sup_U \prod_{i=1}^n U(x_i|y_i). \label{eq-equiv-reps-u}
    \end{align}
    To see this, note that for any $U \gg \hat{P}_{\bs{x}|\bs{y}}$,
    \begin{align}
        \prod_{i=1}^n \hat{P}_{\bs{x}|\bs{y}}(x_i|y_i) 
        &= \left(\prod_{i=1}^n U(x_i|y_i) \right) \!\! \left(\prod_{i=1}^n \frac{\hat{P}_{\bs{x}|\bs{y}}(x_i|y_i)}{U(x_i|y_i)} \right). \label{eq-mmi-eq-1}
    \end{align}
    Now, observe that the second term can be expressed as
    \begin{align}
        \prod_{i=1}^n \frac{\hat{P}_{\bs{x}|\bs{y}}(x_i|y_i)}{U(x_i|y_i)}
        &= \prod_{(a, b)} \left[ \frac{\hat{P}_{\bs{x}|\bs{y}}(a|b)}{U(a|b)}\right]^{n\hat{P}_{\bs{xy}}(a, b)}\\
        &= \exp\left(nD(\hat{P}_{\bs{x}|\bs{y}} \| U | \hat{P}_{\bs{y}})\right),
    \end{align}
    where $D(\cdot || \cdot | \cdot)$ denotes the conditional relative entropy, as defined in \cite[Eq.~2.4]{csiszar2011information}.
    Hence, we have
    \begin{align}
        \prod_{i=1}^n \hat{P}_{\bs{x}|\bs{y}}(x_i|y_i) 
        &= \left(\prod_{i=1}^n U(x_i | y_i)\right)\, e^{nD(\hat{P}_{\bs{x}|\bs{y}} \| U | \hat{P}_{\bs{y}})}. \label{eq-mmi-eq-4}
    \end{align}
    Given that the relative entropy is always non-negative and \eqref{eq-mmi-eq-4} holds for any $U$, we can take the supremum to obtain
    \begin{align}
        \hat{P}_{\bs{x}|\bs{y}}(\bs{x}|\bs{y}) &\geq \sup_U \prod_{i=1}^n U(x_i|y_i),
    \end{align}
    with equality achieved when $U=\hat{P}_{\bs{x}|\bs{y}}$. 
    Recall that $\hat{P}_{\bs{x}} = Q_X$ under constant-composition coding. 
    Thus,
    \begin{align}
        \bar{p}_e^{\mathrm{\,\, mmi}}
        &\, \dot{\leq} \,\,\mathbb{E}\left[M ^{\rho} \, \left( \frac{ \prod_{i=1}^n Q_X(X_i)  }{\sup_U \prod_{i=1}^n U(X_i|Y_i)}\right)^{\rho}\right]\\
        &\, \dot{\leq} \,\, \inf_U\mathbb{E}\left[ \mathrm{e}^{n\rho R} \, \left( \frac{ \prod_{i=1}^n Q_X(X_i)  }{\prod_{i=1}^n U(X_i|Y_i)}\right)^{\rho}\right], \label{eq-constant-composition-22}
    \end{align}
    where the dot in \eqref{eq-constant-composition-22} arises because $M \doteq \exp(nR)$. The upper bound is obtained by moving the minimization outside the expectation. 
    To see this, observe that for any $\bar{\bs{x}}\in \mcl{X}^n$, $\bar{\bs{y}} \in \mcl{Y}^n$, and $\bar{U}$, we have
     \begin{align}
     	\inf_U \left(\frac{Q^n(\bs{\bar{x}})}{U^n(\bs{\bar{x}}|\bs{\bar{y}}) } \right)^{\rho}    
	&\leq \left(\frac{Q^n(\bs{\bar{x}})}{\bar{U}^n(\bs{\bar{x}}|\bs{\bar{y}}) } \right)^{\rho} .   
     \end{align}
     The result then follows by taking the expectation with respect to $(\bar{\bs{x}}, \bar{\bs{y}})$ and subsequently minimizing over $\bar{U}$ on the RHS.
    Taking advantage of constant-composition coding, we simplify the expectation as follows
     \begin{align}
        \begin{split}
            &\mathbb{E}\left[\left( \frac{ \prod_{i=1}^n Q_X(X_i)  }{\prod_{i=1}^n U(X_i|Y_i)}\right)^{\rho}\right]\\
            &=\!\!\!\!\!\! \sum_{\bs{x} \in \mcl{T}^n(Q_X)} \!\! \frac{1}{\left|\mcl{T}^n(Q_X)\right|} \sum_{\bs{y}} \prod_{i=1}^n W({y}_i | {x}_i) \left(\frac{ Q_X(x_i)}{U(x_i | y_i)}\right)^\rho  
        \end{split}\label{eq-constant-composition-2}  \\
        &=\!\!\!\!\!\! \sum_{\bs{x} \in \mcl{T}^n(Q_X)} \!\! \frac{1}{\left|\mcl{T}^n(Q_X)\right|} \prod_{i=1}^n \sum_y W(y | x_i)\left(\frac{Q_X(x_i)}{U(x_i | y)}\right)^\rho\\
        &=\!\!\!\!\!\! \sum_{\bs{x} \in \mcl{T}^n(Q_X)} \!\! \frac{1}{\left|\mcl{T}^n(Q_X)\right|} \! \prod_{a \in \mcl{X}}\!\left[ \sum_y W(y | a)\!\left(\frac{Q_X(a)}{U(a | y)}\right)^\rho\!\right]^{nQ_X(a)} \myeqnum \label{eq-constant-composition-1}  \\
        &= \prod_{x \in \mcl{X}} \left[\sum_y W(y | x)\left(\frac{Q_X(x)}{U(x | y)}\right)^\rho\right]^{nQ_X(x)}, \label{eq-constant-composition}
    \end{align}
    where \eqref{eq-constant-composition} holds because the product term in \eqref{eq-constant-composition-1} is identical for all $\bs{x} \in \mcl{T}^n(Q_X)$.
    Accordingly, we have for any $\rho \in [0, 1]$,
    \begin{align}
        \bar{p}_e^{\,\,\mathrm{mmi}} \,\, \dot{\leq}\,\, \exp\left(-n\left[E_0^{\mathrm{mmi}}(Q_X, \rho) - \rho R \right] \right),
    \end{align}
    where
    \begin{align}
            &E_0^{\mathrm{mmi}}(Q_X, \rho) \notag\\
            &\hspace{0.15cm}= \sup_U -\sum_x Q_X(x) \log \sum_y W(y | x)\left(\frac{Q_X(x)}{U(x | y)}\right)^\rho.\label{eq-move-outer}
    \end{align}
    For any non-negative function $f$ defined on $\mcl{X}$, we have
    $\mathbb{E}[\log f(X)] = \inf_{r(.)} \log \mathbb{E}\left[\frac{f(X)}{e^{r(X)-\phi_r}}\right]$, where $\phi_r = \mathbb{E}\left[r(X)\right]$
     \cite[Section~2.4.2]{scarlett2014}.
    With this approach, the outer expectation in \eqref{eq-move-outer} can be moved inside the logarithm, resulting in
    \begin{align}
        \begin{split}
            &E_0^{\mathrm{mmi}}(Q_X, \rho) \\
            &\hspace{0.15cm}= \sup_U \sup_{r(.)} - \log \sum_{(x,y)} Q_X(x) W(y | x)\left(\frac{Q_X(x)}{U(x | y)e^{r(x)-\phi_r}}\right)^\rho \label{eq-min-u-1}
        \end{split}\\
        &\hspace{0.15cm}= \sup_{r(.)} - \log \inf_U \sum_{(x,y)} Q_X(x) W(y | x)\left(\frac{Q_X(x)}{U(x | y)e^{r(x)-\phi_r}}\right)^\rho. \label{eq-min-u}
    \end{align}
    To solve the optimization over $U$ in \eqref{eq-min-u}, we move the minimization inside the summation, yielding
    \begin{align}
        \begin{split}
            &\inf_U \sum_{(x,y)} Q_X(x) W(y | x)\left(\frac{Q_X(x)}{U(x | y)e^{r(x)-\phi_r}}\right)^\rho\\
            &\hspace{0.1cm}\geq \sum_y \inf_{U(.|y)} \sum_x Q_X(x) W(y | x)\left(\frac{Q_X(x)}{U(x | y)e^{r(x)-\phi_r}}\right)^\rho.  \label{eq-min-u-2}
        \end{split}
    \end{align}
    The expression is convex in $U$, allowing us to apply  \cite[Theorem 4.4.1]{gallager}. 
    This theorem guarantees that, for $x \in \mcl{X}$ and some $\lambda$, the optimal $U^\star(.|y)$ satisfies
    \begin{align}
        Q_X(x) W(y | x)\left(\frac{Q_X(x)}{e^{r(x)-\phi_r}}\right)^\rho U^{\star}(x | y)^{-(1+\rho)}=\lambda,
    \end{align}
    resulting in the optimal solution
    \begin{align}
        U^{\star}(x | y)=\frac{Q_X(x) W(y | x)^{\frac{1}{1+\rho}} e^{\frac{-\rho}{1+\rho}\left[r(x)-\phi_r\right]}}{\sum_{\tilde{x}} Q_X(\tilde{x}) W(y | \tilde{x})^{\frac{1}{1+\rho}} e^{\frac{-\rho}{1+\rho}\left[r(\tilde{x})-\phi_r\right]}}.
    \end{align}
    Note that $U^{\star}$ is indeed the solution to the optimization problem in \eqref{eq-min-u-1}. Setting $U = U^{\star}$ in \eqref{eq-min-u-1} provides a valid lower bound, as $U^{\star}$ is potentially sub-optimal. However, as demonstrated, $U^{\star}$ also satisfies the upper bound. Since the lower and upper bounds are identical, this confirms that $U^{\star}$ is the optimal solution. 
    Thus, by substituting $U^{\star}$, we obtain
    \begin{align}
        \begin{split}
            &\inf_U \sum_{(x,y)} Q_X(x) W(y | x)\left(\frac{Q_X(x)}{U(x | y)e^{r(x)-\phi_r}}\right)^\rho\\\
            &\hspace{0.5cm}= \sum_y \left[\sum_x Q_X(x) W(y|x)^{\frac{1}{1+\rho}} e^{\frac{-\rho}{1+\rho}[r(x)-\phi_r]}\right]^{1+\rho}.
        \end{split}
    \end{align}
    By redefining $r(x) := \frac{-\rho}{1+\rho}r(x)$, the optimization over $r(.)$ remains unchanged. Consequently, we can express $E_0^{\mathrm{mmi}}(Q_X, \rho)$ as
    \begin{align}
        \begin{split}
            &E_0^{\mathrm{mmi}}(Q_X, \rho) \\
            &\hspace{0.15cm}= \sup_{r(.)} -\log \sum_y \left[\sum_x Q_X(x) W(y|x)^{\frac{1}{1+\rho}} e^{r(x)-\phi_r}\right]^{1+\rho},
        \end{split}
    \end{align}
    which is equal to $E_0^{\mathrm{ml}}(Q_X, \rho)$, as given in \cite[Eq. (53)]{6763080}. 
\end{proof}

\section{Joint Source Channel Coding}

We now consider the transmission of non-equiprobable messages with distribution $P^k(\bs{v}) = \prod_{i=1}^k P_V(v_i)$, where $\bs{v} = (v_1, \ldots, v_k) \in \mcl{V}^k$ is the source message, and $\mcl{V}$ is a finite discrete alphabet. 
The source message set $\mcl{V}^k$ is partitioned into $N_k$ disjoint subsets (or classes) $\mcl{A}_k^{(i)}$, $i = 1, \ldots, N_k$, such that $\bigcup_{i=1}^{N_k} \mcl{A}_k^{(i)} = \mcl{V}^k$, where $N_k$ can grow sub-exponentially with $k$. We assume that each class $\mcl{A}_k^{(i)}$, $i = 1, \ldots, N_k$ includes one or more full type classes.
For each source message $\bs{v}$ in $\mcl{A}_k^{(i)}$, the codewords $\bs{x}(\bs{v}) \in \mcl{X}^n$ are generated independently and uniformly from the type class $\mcl{T}^n(Q_i)$, where the size of the type class must satisfy $|\mcl{T}^n(Q_i)| \geq |\mcl{A}_k^{(i)}|$ for all $i = 1, \ldots, N_k$.
%
The work in \cite[Theorem 1]{6803047} derives the MAP random coding error exponents for an arbitrary partition of $\mcl{V}^k$ and recovers Csisz\'ar's exponent \cite{csiszar1980}.
We next show that the exponent of \cite[Theorem 1]{6803047} can be achieved with the generalized MMI decoder, introduced by Csiszár in \cite{csiszar1980}.
Given the channel output $\bs{y}$, the decoder selects the source message $\hat{\bs{v}}$ as follows
\begin{align}
    \hat{\bs{v}} = \underset{\bs{v}}{\operatorname{argmax}} \,\, I(\hat{P}_{\bs{x}(\bs{v})\bs{y}}) - {tH(\hat{P}_{\bs{v}})},
    \label{eq:mmi_gen_jscc}
\end{align}
where $t\triangleq \frac k n$. 
Similarly to the MMI decoder in channel coding, this decoder does not require knowledge of the source or channel distribution.
Following similar steps as in \eqref{eq-equality-py-0}-\eqref{eq-equality-py-last}, it can be shown that \eqref{eq:mmi_gen_jscc} is equivalent to
\begin{align}
    \hat{\bs{v}} = \underset{\bs{v}}{\operatorname{argmax}} \left(\prod_{\ell=1}^k \hat{P}_{\bs{v}}(v_{\ell})\right) \left(\prod_{\ell=1}^n \hat{P}_{\bs{y}|\bs{x}(\bs{v})}({y}_{\ell}|{x}_{\bs{v},\ell})\right).
\end{align}
\begin{theorem}
    For a given partition $\mcl{A}_k^{(i)}$, $i = 1, \ldots, N_k$, and associated random-coding distributions $Q_i$, the average probability of error satisfies
    \begin{align}
        \begin{split}
            \bar{p}_e^{\mathrm{\,\,mmi}}
            &\, \dot{\leq}\,  \sum_{i=1}^{N_k} \exp \left(-\!\!\max _{\rho_i \in[0,1]}\left\{n E_0^{\mathrm{ml}}(Q_i, \rho_i)-E_{\mathrm{s}}^{(i)}(\rho_i, P^k)\right\}\right)
        \end{split}
    \end{align}
    where
    $E_{\mathrm{s}}^{(i)}(\rho, P^k) \triangleq \log \left(\sum_{\bs{v} \in \mcl{A}_k^{(i)}} P^k(\bs{v})^{\frac{1}{1+\rho}}\right)^{1+\rho}$.
\end{theorem}

\begin{proof}
    We begin with the extended RCU bound for joint source-channel coding, given in Eq. 26 of \cite{6803047}, and adapt it to the generalized MMI decoder as follows
    \begin{align}
        \bar{p}_e^{\mathrm{\,\,mmi}}
        &\leq \sum_{i=1}^{N_k} \bar{\varepsilon}_i,
    \end{align}
    where
    \begin{align}
        \begin{split}
            &\bar{\varepsilon}_i = \sum_{\bs{v} \in \mcl{A}_k^{(i)}} P^k(\bs{v}) \sum_{(\bs{x}, \bs{y})} P_{\bs{X}}^{(i)}(\bs{x}) W^n(\bs{y} | \bs{x})\\
            &\hspace{0.4cm}\times \min \Bigg\{1, \sum_{j=1}^{N_k} \sum_{\bar{\bs{v}} \in \mcl{A}_k^{(j)}} \,\, \sum_{\substack{\bar{\bs{x}}: \hat{P}_{\bs{\bar{v}}}(\bar{\bs{v}}) \hat{P}_{\bs{y} | \bs{\bar{x}}}(\bs{y} | \bar{\bs{x}}) \\ \, \geq \hat{P}_{\bs{v}}(\bs{v}) \hat{P}_{\bs{y} | \bs{x}}(\bs{y} | \bs{x})}} P_{\bs{X}}^{(j)}(\bar{\bs{x}})\Bigg\},
        \end{split}
    \end{align}
	and $P_{\bs{X}}^{(i)}(\bs{x})$ denotes the constant-composition ensemble distribution, as defined in \eqref{eq-cc-defn} for $Q_i$.
	Next, we use Markov's inequality, once again without applying tilting, to obtain
    \begin{align}
        \sum_{\substack{\bar{\bs{x}}: \hat{P}_{\bs{\bar{v}}}(\bar{\bs{v}}) \hat{P}_{\bs{y} | \bs{\bar{x}}}(\bs{y} | \bar{\bs{x}}) \\  \geq \hat{P}_{\bs{v}}(\bs{v}) \hat{P}_{\bs{y} | \bs{x}}(\bs{y} | \bs{x})}} \!\!\!\!\!\!\!\!\!\!\! P_{\bs{X}}^{(j)}(\bar{\bs{x}}) 
        &\leq \sum_{\bar{\bs{x}}} P_{\bs{X}}^{(j)}(\bar{\bs{x}}) \left( \frac{\hat{P}_{\bs{\bar{v}}}(\bar{\bs{v}}) \hat{P}_{\bs{y} | \bs{\bar{x}}}(\bs{y} | \bar{\bs{x}})}{\hat{P}_{\bs{v}}(\bs{v}) \hat{P}_{\bs{y} | \bs{x}}(\bs{y} | \bs{x})}\right).
    \end{align}
    We can again follow similar steps as in \eqref{eq-pol0}-\eqref{eq-pol2}, using  \cite[Lemma~2.1]{poltyrev}, to obtain
    \begin{align}
        \sum_{\bar{\bs{x}}} P_{\bs{X}}^{(j)}(\bar{\bs{x}}) \hat{P}_{\bs{y} | \bs{\bar{x}}}(\bs{y} | \bar{\bs{x}}) \,\, \dot{\leq} \,\, \hat{P}_{\bs{y}} (\bs{y}).
    \end{align}
    Thus, we have
    \begin{align}
        \begin{split}
            &\bar{\varepsilon}_i \,\, \dot{\leq} \,\, \sum_{\bs{v} \in \mcl{A}_k^{(i)}} P^k(\bs{v}) \sum_{(\bs{x}, \bs{y})} P_{\bs{X}}^{(i)}(\bs{x}) W^n(\bs{y} | \bs{x})\\
            &\hspace{0.35cm}\times \min \Bigg\{1, \Bigg(\sum_{j=1}^{N_k} \sum_{\bar{\bs{v}} \in \mcl{A}_k^{(j)}} \hat{P}_{\bs{\bar{v}}}(\bar{\bs{v}}) \!\! \Bigg)  \frac{\hat{P}_{\bs{y}} (\bs{y})}{\hat{P}_{\bs{v}}(\bs{v}) \hat{P}_{\bs{y} | \bs{x}}(\bs{y} | \bs{x})} \Bigg\}.
        \end{split}
    \end{align}
    Let $\Lambda_i$ denote the set of all possible types of source sequences in $\mcl{A}_k^{(i)}$ (i.e., $\Lambda_i \subset \mcl{P}_k(\mcl{V})$). 
    Consequently, the summation in the minimum function simplifies to
    \begin{align}
        \sum_{j=1}^{N_k} \sum_{\bar{\bs{v}} \in \mcl{A}_k^{(j)}} \hat{P}_{\bs{\bar{v}}}(\bar{\bs{v}})
        &= \sum_{j=1}^{N_k} \sum_{\bar{P} \in \Lambda_j} \sum_{\bar{\bs{v}} \in \mcl{T}^k(\bar{P})} \hat{P}_{\bs{\bar{v}}}(\bar{\bs{v}})\\
        &= \sum_{j=1}^{N_k} \sum_{\bar{P} \in \Lambda_j} \sum_{\bar{\bs{v}} \in \mcl{T}^k(\bar{P})} \exp(-kH(\bar{P}))\\
        &\doteq \sum_{j=1}^{N_k} \sum_{\bar{P} \in \Lambda_j} 1 \label{eq-appear-one-2}\\
        &\doteq 1,
    \end{align}
    where the $1$ in \eqref{eq-appear-one-2} arises from the fact that $|\mcl{T}^k(\bar{P})| \doteq \exp(kH(\bar{P}))$, and the final dot-equality holds because the number of possible types in $\mcl{A}_k^{(j)}$ and $N_k$ both grow sub-exponentially with $k$. Hence, similarly to \eqref{eq:pxy_div} we have
    \begin{align}
        \begin{split}
            &\bar{\varepsilon}_i 
            = \sum_{\bs{v} \in \mcl{A}_k^{(i)}} P^k(\bs{v}) \sum_{(\bs{x}, \bs{y})} P_{\bs{X}}^{(i)}(\bs{x}) W^n(\bs{y} | \bs{x})\\
            &\hspace{1.25cm}\times \min \left\{1, \,\, \frac{\hat{P}_{\bs{x}}(\bs{x})}{\hat{P}_{\bs{v}}(\bs{v}) \hat{P}_{\bs{x}|\bs{y} }(\bs{x}|\bs{y})}\right\}.
        \end{split}
    \end{align}
    Observe that $\hat{P}_{\bs{x}} = Q_i$, and recall the inequality $\min\{1, x\} \leq x^{\rho}$ for $x \geq 0$ and $\rho \in [0, 1]$. Hence, for any $\rho_i \in [0, 1]$,
    \begin{align}
        \begin{split}
            &\bar{\varepsilon}_i \, \, \dot{\leq} \!\!\! \sum_{\bs{v} \in \mcl{A}_k^{(i)}}\!\! P^k(\bs{v}) \!\!\!\!\!\!\!\!\sum_{(\bs{x} \in \mcl{T}^n(Q_i), \bs{y})} \frac{ W^n(\bs{y} | \bs{x})}{| \mcl{T}^n(Q_i) |} \left(\frac{Q_i^n(\bs{x})}{\hat{P}_{\bs{v}}(\bs{v}) \hat{P}_{\bs{x}|\bs{y} }(\bs{x}|\bs{y})}\right)^{\rho_i}.
        \end{split}
    \end{align}
    Recall the equivalent representation of $\hat{P}_{\bs{x}|\bs{y}}(\bs{x}|\bs{y})$ given in \eqref{eq-equiv-reps-u}. 
    Substituting this representation, we can further upper bound $\bar{\varepsilon}_i$ as follows
    \begin{align}
        \begin{split}
            \bar{\varepsilon}_i \,\,&\dot{\leq} \,\, \Bigg(\sum_{\bs{v} \in \mcl{A}_k^{(i)}} \frac{P^k(\bs{v})}{\hat{P}_{\bs{v}}(\bs{v})^{\rho_i}}\Bigg)\\
            &\hspace{0.5cm}\times \Bigg(\inf_U \!\!\!\! \sum_{(\bs{x} \in \mcl{T}^n(Q_i), \bs{y})} \frac{W^n(\bs{y} | \bs{x})}{\left|\mcl{T}^n\left(Q_i\right)\right|}\left(\frac{Q_i^n(\bs{x})}{U^n(\bs{x} | \bs{y})}\right)^{\rho_i}\Bigg).
        \end{split}
        \label{eq:jscc_2mid}
    \end{align}
    Observe that the second term in \eqref{eq:jscc_2mid} is identical to the expression in \eqref{eq-constant-composition-2}, and thus,
    \begin{align}
        \bar{\varepsilon}_i \,\, \dot{\leq} \, \Bigg(\sum_{\bs{v} \in \mcl{A}_k^{(i)}} \frac{P^k(\bs{v})}{\hat{P}_{\bs{v}}(\bs{v})^{\rho_i}}\Bigg) \, \exp(-nE_0^{\mathrm{ml}}(Q_i, \rho_i)).
        \label{eq:jscc_mid}
    \end{align}
    It remains to be shown that the first term in \eqref{eq:jscc_mid} is upper bounded by $\exp\left(E_{\mathrm{s}}^{(i)}(\rho_i, P^k)\right)$. To this end, we have
    \begin{align}
        \sum_{\bs{v} \in \mcl{A}_k^{(i)}} \frac{P^k(\bs{v})}{\hat{P}_{\bs{v}}(\bs{v})^{\rho_i}} 
        &=\sum_{\bar{P} \in \Lambda_i} \sum_{\bs{v} \in \mcl{T}^k(\bar{P})}  \frac{P^k(\bs{v})}{\bar{P}^k(\bs{v})^{\rho_i}}. \label{eq-source-type-simple}
    \end{align}
    Observe that for $\bs{v} \in \mcl{T}^k(\bar{P})$, we have $\frac{P^k(\bs{v})}{\bar{P}^k(\bs{v})^{\rho_i}} = \exp\left(k\sum_{b \in \mcl{V}} \bar{P}(b)\log\frac{P_V(b)}{\bar{P}(b)^{\rho_i}}\right)$, and $\left|\mcl{T}^k(\bar{P})\right| \doteq \exp(kH(\bar{P}))$. 
    Consequently, \eqref{eq-source-type-simple} simplifies to
    \begin{align}
        \sum_{\bs{v} \in \mcl{A}_k^{(i)}}  \! \frac{P^k(\bs{v})}{\hat{P}_{\bs{v}}(\bs{v})^{\rho_i}} 
        & \doteq \!\!\! \sum_{\bar{P} \in \Lambda_i} \!\! \exp\!\left(k\sum_{b\in \mcl{V}}\bar{P}(b)\log\frac{P_V(b)}{\bar{P}(b)^{1+\rho_i}}\right).\label{eq-source-type-simple-1}
    \end{align}
    The exponential term on the RHS of \eqref{eq-source-type-simple-1} can be rewritten as
    \begin{align}
        \begin{split}
            &\exp\left(k\sum_{b\in \mcl{V}}\bar{P}(b)\log\frac{P_V(b)}{\bar{P}(b)^{1+\rho_i}}\right)\\
            &\hspace{1cm}=\left[\exp\left(k \sum_{b\in \mcl{V}}\bar{P}(b)\log\frac{P_V(b)^{\frac{1}{1+\rho_i}}}{\bar{P}(b)}\right)\right]^{1+\rho_i}
        \end{split}\\
        &\hspace{1cm}=\Bigg[\sum_{\bs{v} \in \mcl{T}^k(\bar{P})} P^k(\bs{v})^{\frac{1}{1+\rho_i}}\Bigg]^{1+\rho_i}.
    \end{align}
    Substituting this back into \eqref{eq-source-type-simple}, we have
    \begin{align}
        \sum_{\bs{v} \in \mcl{A}_k^{(i)}} \frac{P^k(\bs{v})}{\hat{P}_{\bs{v}}(\bs{v})^{\rho_i}} 
        &= \sum_{\bar{P}\in \Lambda_i} \Bigg[\sum_{\bs{v} \in \mcl{T}^k(\bar{P})} \!\!\!\! P^k(\bs{v})^{\frac{1}{1+\rho_i}} \Bigg]^{1+\rho_i}\\
        &\leq \Bigg[\sum_{\bar{P}\in \Lambda_i}  \sum_{\bs{v} \in \mcl{T}^k(\bar{P})} \!\!\!\! P^k(\bs{v})^{\frac{1}{1+\rho_i}} \Bigg]^{1+\rho_i} \label{eq-gal-holder-f} \\
        &= \Bigg[ \sum_{\bs{v} \in \mcl{A}_k^{(i)}} P^k(\bs{v})^{\frac{1}{1+\rho_i}} \Bigg]^{1+\rho_i},
    \end{align}
    where in \eqref{eq-gal-holder-f}, we used $\sum_i a_i^r \leq (\sum_i a_i)^r$ for $r \geq 1$ \cite[Problem~4.15.f]{gallager}. 
    By combining these results and minimizing over $\rho_i$, we obtain
    \begin{align}
        \bar{\varepsilon}_i  \,\, \dot{\leq} \, \exp \! \left(-\!\!\max _{\rho_i \in[0,1]}\left\{n E_0^{\mathrm{ml}}(Q_i, \rho_i)-E_{\mathrm{s}}^{(i)}(\rho_i, P^k)\right\}\right),
    \end{align}
    thus concluding the proof.
\end{proof}

\bibliographystyle{IEEEtran}
\bibliography{refs}

\end{document}